\documentclass[a4paper,fleqn,usenatbib]{mnras}

\usepackage[T1]{fontenc}
\usepackage{ae,aecompl}
\usepackage{xcolor}

\usepackage{graphicx}	
\usepackage{amsmath}	
\usepackage{amssymb}	
\usepackage{pdflscape}	
\usepackage[english]{babel}

\newcommand{\msol}{\mathrm{M}_{\rm \odot}}
\newcommand{\rsol}{\mathrm{R}_{\rm \odot}}
\newcommand{\mjup}{\mathrm{M}_{\rm Jup}}
\newcommand{\rjup}{\mathrm{R}_{\rm Jup}}
\newcommand{\mearth}{\mathrm{M}_{\rm \oplus}}
\newcommand{\rearth}{\mathrm{R}_{\rm \oplus}}

\title[Exomoon detection in exoplanet phase curves]{On The Feasibility of Exomoon Detection Via Exoplanet Phase Curve Spectral Contrast}

\author[D. H. Forgan]
{D.~H.~Forgan$^{1,2}$\thanks{Contact e-mail: \href{mailto:dhf3@st-andrews.ac.uk}{dhf3@st-andrews.ac.uk}}
\vspace{0.2cm} \\
$^{1}$SUPA, School of Physics \& Astronomy, University of St Andrews, North Haugh, St Andrews, Scotland, KY16 9SS, UK\\
$^{2}$St Andrews Centre for Exoplanet Science}

\date{Accepted XXX. Received YYY; in original form ZZZ}

\pubyear{2016}

\begin{document}
\label{firstpage}
\pagerange{\pageref{firstpage}--\pageref{lastpage}}
\maketitle

\begin{abstract}
An exoplanet-exomoon system presents a superposition of phase curves to observers - the dominant component varies according to the planetary period, and the lesser varies according to both the planetary and the lunar period.  If the spectra of the two bodies differs significantly, then it is likely there are wavelength regimes where the contrast between the moon and planet is significantly larger.  In principle, this effect could be used to isolate periodic oscillations in the combined phase curve.  Being able to detect the exomoon component would allow a characterisation of the exomoon radius, and potentially some crude atmospheric data.    

We run a parameter survey of combined exoplanet-exomoon phase curves, which show that for most sets of planet-moon parameters, the lunar component of the phase curve is undetectable to current state-of-the-art transit observations.  Even with future transit survey missions, measuring the exomoon signal will most likely require photometric precision of 10 parts per million or better.  

The only exception to this is if the moon is strongly tidally heated or in some way self-luminous.  In this case, measurements of the phase curve at wavelengths greater than a few microns can be dominated by the lunar contribution.  Instruments like the James Webb Space Telescope and its successors are needed to make this method feasible.

\end{abstract}




\section{Introduction}
\label{sec:introduction}

While the detection of extrasolar planets (exoplanets) continues apace\footnote{http://exoplanets.eu}, the detection of their satellites, extrasolar moons or exomoons remain undetected.  This is not for want of trying - several teams are attempting to achieve this first detection.  The community has been able to deliver strong upper limits on the sizes of exomoons in their samples (see e.g. \citealt{Kipping2015}), and these constraints are expected to grow tighter in the coming years as the exoplanet transit missions CHEOPS \citep{Simon2015} and PLATO \citep{Hippke2015} come online.

This lack of exomoon detections is also not for want of exomoon detection methods.  The recent literature abounds with indirect and direct techniques.  Some examples include transit timing and duration variations (TTVs/TDVs) \citep{moon_detect,Simon2007,Kipping2009,Kipping2009a,Lewis2013,Heller2016}, direct transits of exomoons (e.g. \citealt{Brown2001,Pont2007,Dobos2016}), microlensing events \citep{Han2002,Bennett2013}, mutual eclipses of directly imaged planet-moon systems \citep{Cabrera2007} or mutual eclipses during a stellar transit \citep{Sato2009a}.  Averaging of multiple transits may also yield an exomoon signal, either through scatter peak analysis \citep{Simon2012a} or orbital sampling of light curves \citep{Heller2016a}.  In the radio, emission from giant planets may be modulated by the presence of moons within or near the magnetosphere \citep{Noyola2014}, or indeed by moon-induced plasma torii \citep{Ben-Jaffel2014}.


Many of these methods are relatively agnostic to the wavelength of the observation.  Some very recent proposals for detection methods rely on differences in the spectra of the planet and moon.  For example, the spectro-astrometric detection method proposed by \citet{Agol2015} attends to directly-imaged exoplanet-exomoon systems.  Direct imaging lacks the spatial resolution to separately image the moon - however, comparing observations at two wavelengths can identify a shift in centroid, if one wavelength happens to coincide with a regime where the planet is faint and the moon is bright.  For example, an icy moon is generally quite reflective across a variety of wavelengths, but a giant planet may contain significant (and relatively wide) absorption features (such as the methane feature at around 1.4 $\micron$).
 

We consider a new, but related spectral detection method.  Photometric observations of transiting exoplanets across a full orbital period obtain primary and secondary transits (where the planet eclipses the star and the star eclipses the planet respectively), and the phase curve, a smoothly varying component which increases as the planet's dayside comes in and out of view \citep[see e.g.][]{Winn2011,Madhusudhan2012}.  The phase curve's period is that of the planet's orbit around the star.  Any exomoons present will also make contributions to the phase curve.  These contributions will have multiple components, whose periods are equal to the planetary orbital period and the lunar orbital period (depending on the radiation budgets of both bodies).

In principle, if there are regions of the spectrum where the contrast between the moon and planet is high, the exomoon's phase curve may be isolated by subtracting the modelled exoplanet phase curve, if it is measured at multiple wavelengths.  In essence, we are searching for a periodicity in the total phase curve that can only be explained by the presence of an exomoon.  

We investigate the efficacy of this detection method, which we christen \emph{phase curve spectral contrast} (PCSC).  This concept builds on the work of \citet{Robinson2011}\footnote{In fact, the genesis of this idea may be traced as far back as \citet{Williams2004} and \citet{Selsis2004}}, who considered the time-dependence of the Earth-Moon's combined spectrum, and advocated a phase differencing between e.g. full phase and quadrature.  The technique discussed here is similar, but we focus less on the Earth-Moon system, and consider a more generalised approach, with a consequently larger parameter space.  We also attempt to use the entire phase curve, rather than two instantaneous measurements.  

\citet{Moskovitz2009} also presented a detailed study of the Moon's effect on the Earth's infrared phase curve using both a 1D energy balance model and a 3D general circulation climate model, but these results were focused on a single wavelength of observation (see also \citealt{Gomez-Leal2012}'s use of real atmospheric data in a similar study).  Multi-frequency observations will be crucial for detections, especially in the absence of detailed knowledge of the atmospheric properties of each body.


This paper is structured as follows: in section \ref{sec:analytics} we motivate this detection method by considering how spectral contrast can assist in extracting exomoon's contribution to the phase curve, and we describe code we have written to investigate the phase curves produced by exoplanet-exomoon systems; in section \ref{sec:results} we give examples of the combined phase curves produced in a series of limiting cases, and consider the exomoon parameter space that could be probed by this detection method given present and future transit surveys; finally in section \ref{sec:conclusions}, we summarise the work.

\section{Analytical Description}
\label{sec:analytics}

\noindent We can write the flux received from an exoplanet, $F_p$ as the sum of reflected stellar radiation and the planet's thermal radiation:

\begin{equation}
F_p = F_* \alpha_p \left(\frac{R_p}{a_{p*}}\right)^2 + F_{p,t},
\end{equation}

\noindent where $F_*$ is the stellar flux, $\alpha_p$ is the planet's albedo, $R_p$ is the planetary radius, $a_{p*}$ is the semimajor axis of the planet's orbit, and $F_{p,t}$ is the planet's thermal flux.  We can write a similar equation for the lunar flux, $F_s$:

\begin{equation}
F_s = F_* \alpha_s \left(\frac{R_s}{a_{s*}}\right)^2 + F_p \alpha_s\left(\frac{R_s}{a_{ps}}\right)^2 + F_{s,t}.
\end{equation}

\noindent The lunar flux now includes reflective contributions from the star and the planet respectively.  In the limit that $a_{p*}>>a_{ps}$, we can approximate $a_{s*} = a_{p*}$.  Expanding $F_p$ and rearranging gives:

\begin{equation}
F_s = F_* \alpha_s \left(\frac{R_s}{a_{p*}}\right)^2\left( 1 + \alpha_p\left(\frac{R_p}{a_{ps}}\right)^2\right)  +  F_{p,t} \alpha_s\left(\frac{R_s}{a_{ps}}\right)^2  + F_{s,t},
\end{equation}

\noindent and thus we can write the contrast ratio

\begin{equation}
\frac{F_s}{F_p} = \frac{F_* \alpha_s \left(\frac{R_s}{a_{p*}}\right)^2\left( 1 + \alpha_p\left(\frac{R_p}{a_{ps}}\right)^2\right)  +  \alpha_s\left(\frac{R_s}{a_{ps}}\right)^2 F_{p,t} + F_{s,t}}{F_* \alpha_p \left(\frac{R_p}{a_{p*}}\right)^2 + F_{p,t}}.
\end{equation}

\noindent To clarify matters, let us consider some limiting cases.  Firstly, in the limit that the thermal contribution of both bodies is negligible, we obtain

\begin{equation}
\frac{F_s}{F_p} = \frac{\alpha_s}{\alpha_p}\left(\frac{R_s}{R_p}\right)^2 + \alpha_s\left(\frac{R_s}{a_{ps}}\right)^2 \label{eq:reflective_regime}
\end{equation}

\noindent The right hand term is likely to be negligible, so we then derive the following reflectivity contrast condition for the moon/planet flux contrast to be significant:

\begin{equation}
\frac{\alpha_s}{\alpha_p} >>\left(\frac{R_p}{R_s}\right)^2.
\end{equation}

This defines our detection limit for non-luminous planets and moons.  In the limit that thermal emission dominates, the flux contrast ratio becomes:

\begin{equation}
\frac{F_s}{F_p} = \alpha_s \left(\frac{R_s}{a_{ps}}\right)^2 + \frac{F_{s,t}}{F_{p,t}} \label{eq:thermal_regime}.
\end{equation}

In this limit, we must rely on the moon being significantly more luminous than the planet (perhaps due to tidal heating effects, cf \citealt{Peters2013,Dobos2015}).

So what exomoons might contrast methods be amenable to? It is interesting to note that there is no direct relation to $a_{p*}$ in either of our detection limits.  For example, a transiting planet at large semimajor axis with a young, hot moon may be amenable to these methods, or an extremely low-albedo planet (although the planet's own phase curve, and subsequent spectral contrast, will of course be affected by distance to the star).  We will investigate the detailed parameter space in the following sections.

\subsection{Computing Combined Exoplanet-Exomoon Phase Curves}

\noindent For simplicity, we will assume that both the exoplanet and the exomoon orbit their host with zero eccentricity, and that both bodies orbit within the $x-y$ plane.  Their orbital motions are simply

\begin{align*}
x_p(t) &= a_{p*} \cos \nu_p(t) \\
y_p(t) &= a_{p*} \sin \nu_p(t) \\ 
x_s(t) &= x_p(t) + a_{ps} \cos \nu_s(t) \\
y_s(t) &= y_p(t) + a_{ps} \sin \nu_s(t),
\end{align*}

\noindent Where we define the orbital longitudes $\nu$ as:

\begin{align*}
\nu_p(t) & = \frac{2\pi t}{P_{planet}} \\
\nu_s(t) & = \frac{2\pi t}{P_{moon}}.
\end{align*}

\noindent The star is fixed at the origin.  An observer $\mathbf{o}$ is placed along the y-axis.  The phase curve of each body depends on the angle $\delta$ between the emitter $\mathbf{e}$ and the observer, from the perspective of the emission recipient $\mathbf{q}$.  We have three emitter-recipient pairs in the system: star-planet, star-moon and planet-moon, and at any given instant we must compute three angles $\delta_i$, where 

\begin{equation}
\cos \delta_i = \frac{(\mathbf{o}-\mathbf{q}).(\mathbf{e}-\mathbf{q})}{\left|\mathbf{o}-\mathbf{q}\right|.\left|\mathbf{e}-\mathbf{q}\right|}.
\end{equation}

\noindent We assume both the exoplanet and exomoon possess isotropically scattering Lambertian surfaces, ensuring both bodies have fully analytic phase curves (the reader can find detailed numerical integrations of phase curves for other scattering laws in \citealt{Madhusudhan2012}).  The phase function $\Phi(\delta)$ in this case is

\begin{equation}
\Phi(\delta)  = \frac{\sin \delta + (\pi - \delta)\cos \delta}{\pi}.
\end{equation}

\noindent We compute the total flux received at time $t$ as follows.  Firstly, we compute the stellar flux reflected by the planet:

\begin{equation}
F_{p*,r} = \alpha_p \left(\frac{R_p}{a_{p*}}\right)^2\Phi(\delta_{p*}(t)),
\end{equation}

\noindent and the same for the moon:

\begin{equation}
F_{s*,r} = \alpha_s \left(\frac{R_s}{a_{s*}}\right)^2\Phi(\delta_{s*}(t)).
\end{equation}

\noindent We then compute the thermal flux from both bodies.  We fix a flux difference between the night and day side of each body (relative to the star) $f_{night}$, and assume the flux from each side is isotropic.  Consequently, we can use a simple cosine function to describe the fraction of dayside visible to the observer \citep{Williams2008}: 

\begin{equation}
\zeta(\delta) =  \frac{1}{2}\left(1 + \cos \delta \right)
\end{equation}

\noindent We can then immediately write the planet's thermal flux

\begin{equation}
F_{p,t}(t) = F_{p,t,0} \left(f_{night,p} + (1-f_{night,p})\zeta(\delta_{p*}(t)) \right).
\end{equation}

\noindent To write the same equation for the satellite, we must know what defines the dayside nightside contrast on the moon, i.e. is the lunar temperature governed by stellar or planetary flux? If the lunar terminator is determined by stellar flux, the resulting phase curve will possess similar time variation to the reflective case. If the lunar terminator is determined by planetary flux, then the lunar's thermal contribution to the phase curve will vary according to the lunar period about the planet, presenting a wholly different signal.  The two cases are (respectively):

\begin{eqnarray}
F_{s,t}(t) & = F_{s,t,0} \left(f_{night,s} + (1-f_{night,s}) \zeta(\delta_{s*}(t))\right) \\
F_{s,t}(t) & = F_{s,t,0} \left(f_{night,s} + (1-f_{night,s}) \zeta(\delta_{sp}(t))\right).
\end{eqnarray}

\noindent Note that we have neglected the moon's thermal flux variation due to the planet's orbit about the star.  Finally, we compute the planetary flux reflected by the moon, which will combine both reflected starlight and thermal planetary radiation:

\begin{multline*}
F_{sp} = \alpha_s \alpha_p \left(\frac{R_s}{a_{s*}}\right)^2  \left(\frac{R_p}{a_{p*}}\right)^2 + \\ 
\alpha_s \left(\frac{R_s}{a_{sp}}\right)^2 F_{p,t,0} \left(f_{night,p} + \frac{1}{2}(1 -f_{night,p})(1- \cos \theta)\right).
\end{multline*}

\noindent In the second term, we account for the moon receiving changing levels of thermal radiation from the planet as it passes between the day side and the night side.  This is described by the angle between the stellar and lunar position vectors, relative to the planet position vector:

\begin{equation}
\cos \theta = \frac{\left(\mathbf{r}_* - \mathbf{r}_p\right).\left(\mathbf{r}_s - \mathbf{r}_p\right)}{\left|\mathbf{r}_* - \mathbf{r}_p\right|\left|\mathbf{r}_s - \mathbf{r}_p\right|} = \cos \nu_p \cos \nu_s - \sin \nu_p \sin \nu_s.
\end{equation}

\subsection{Extracting Lunar Signals from the Phase Curve}

\noindent Essentially, our task is to extract a short period signal from a combined signal containing a variety of periods.  If the planet's mass $M_p$ and semimajor axis $a_{p*}$ are well characterised, then we can focus our search to regions of period space where an exomoon may stably orbit.  

The orbital stability limits for a moon around a planet are defined by the Roche limit at low exomoon semimajor axis, and the Hill Stability criterion at large exomoon semimajor axis.  The inner Roche limit for an exomoon is:

\begin{equation}
R_{Roche} =  R_s \left(\frac{M_p}{M_s}\right)^{1/3}.
\end{equation}

\noindent By application of Kepler's third law, the corresponding lunar orbital period is:

\begin{equation}
P_{Roche} = \frac{4 \pi^2 R^3_s}{G M_s}.
\end{equation}

\noindent The outer stability limit is proportional to the planet's Hill radius $R_H$:

\begin{equation}
R_{stable} = \chi R_H = \chi a_{p*}\left(\frac{M_p}{3M_*}\right)^{1/3},
\end{equation}

\noindent where $\chi=0.3-0.5$ \citep{Domingos2006}.  Again, this can be converted to an orbital period:

\begin{equation}
P_{stable} = \left(\frac{\chi^2}{3}\right)^{1/2} P_{planet} \label{eq:p_stable}.
\end{equation}

\begin{figure}
\begin{center}
\includegraphics[scale=0.4]{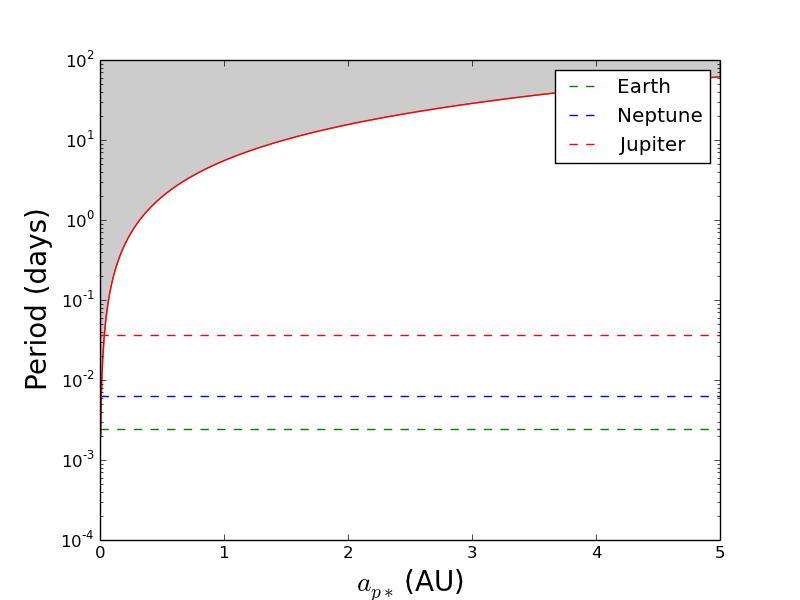}
\end{center}
\caption{The range of permissible orbital periods for a moon orbiting the Earth, Neptune or Jupiter, with a host star of mass $1\msol$.  The dashed lines indicate the Roche limit for the three planets (where we assume the exomoon has $M_s=0.01 \mearth$, i.e. its mass is approximately that of the Moon).  The upper curve indicates the orbital stability limit given by equation (\ref{eq:p_stable}), with $\chi=0.3$.  Shaded regions above the upper curve are orbitally unstable. \label{fig:period_range}}
\end{figure}

\noindent The minimum and maximum permissible orbital periods as a function of planet mass and semimajor axis can be found in Figure \ref{fig:period_range}.

\noindent Two approaches are possible.  If the exomoon contribution in a single band is very strong, we can bandpass filter the combined exoplanet-exomoon phase curve (at a given wavelength) for signals inside the period range $[P_{Roche},P_{stable}]$, where we must assume $M_s$ \emph{a priori} to obtain the lower limit.  We should expect $M_s$ to be significantly lower than $M_p$\footnote{if the planet and moon masses are comparable, the produced transit signals will be markedly different, see \citealt{Lewis2015}}.  Computing the generalised Lomb-Scargle periodogram \citep[GLS, ][]{Zechmeister2009,Mortier2015} of this filtered signal can then identify promising oscillations.  The second approach requires simultaneous measurements of the phase curve in two wavelength bands (one where the planet dominates, and one where the moon dominates).  This can permit a successful subtraction of the exoplanet phase curve to return a periodic residual, which can then be analysed via the GLS periodogram.

\section{Tests \& Discussion}
\label{sec:results}

\subsection{The Highly Reflective Regime}

\noindent As an illustration, we first consider a system containing a $1\msol$ star, a $1 \mjup$ planet orbiting at 0.3 au (orbital period $P_{planet}=0.16$ yr).  We place a Europa-like moon in a circular orbit around the planet, with size $R_s = 0.245 \rearth$, and orbital semimajor axis $a_{ps}=0.0045 $ au.

We consider measurements made in two bands, and hence with two different albedos for each object. The planetary albedo $\alpha_p=[0.3, 10^{-4}]$, and the satellite's albedo $\alpha_s = [10^{-4}, 1.0]$.  These are extreme values, which we select for illustrative purposes only.

\begin{figure*}
\begin{center}$\begin{array}{cc}
\includegraphics[scale=0.4]{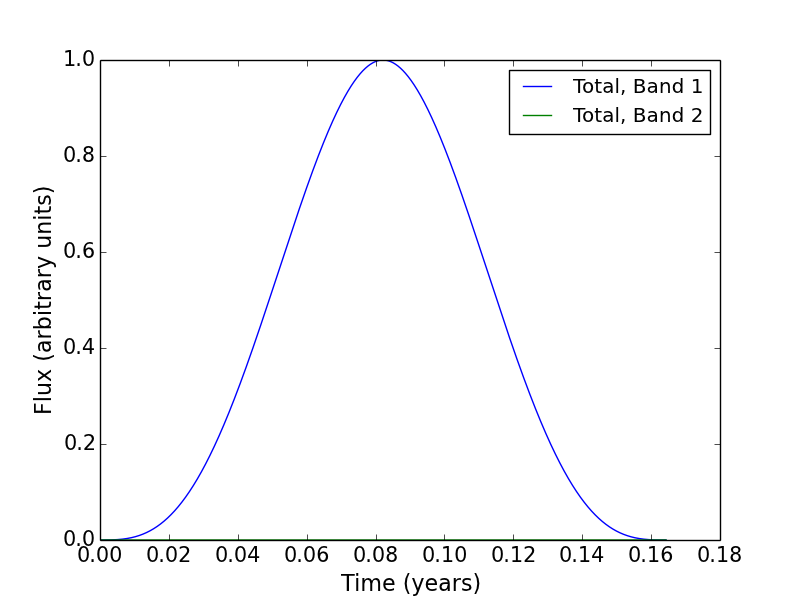} &
\includegraphics[scale=0.4]{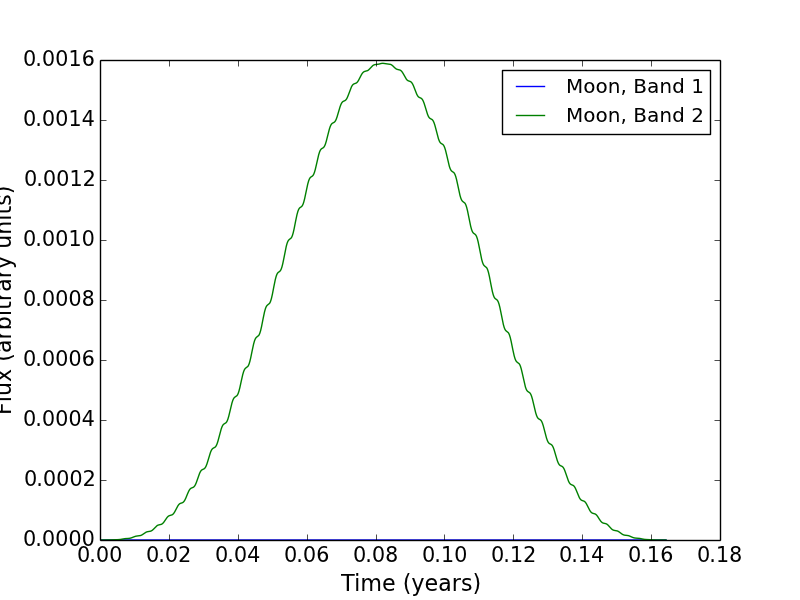} \\
\includegraphics[scale=0.3]{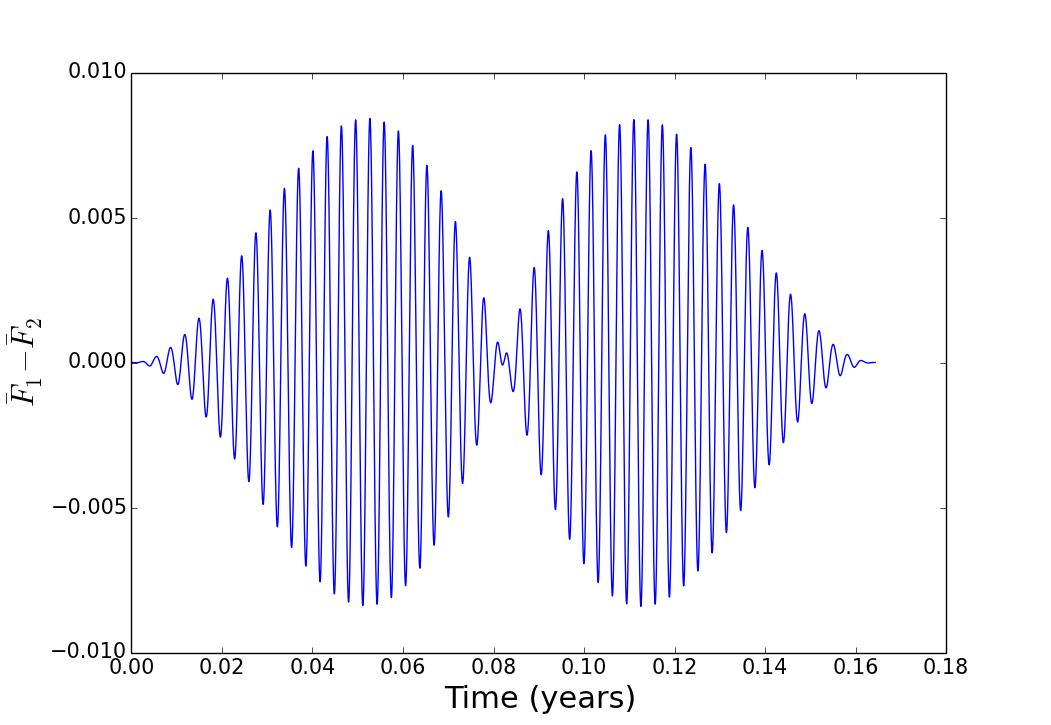}&
\includegraphics[scale=0.4]{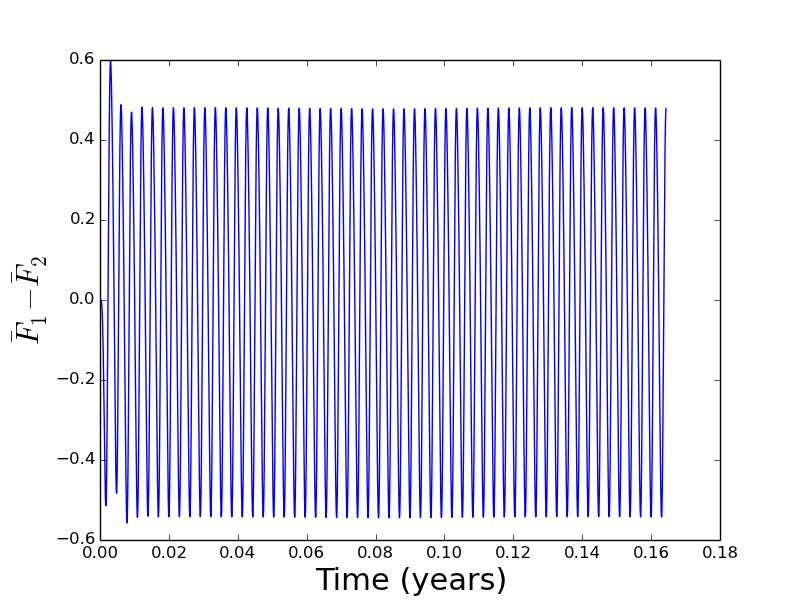} \\
\end{array}$
\end{center}
\caption{Isolating the exomoon contribution to the phase curve in the highly reflective regime.  Top row: The individual contributions to the phase curve from a Jupiter-Europa system placed at $a_{p*}=0.3$ au from a solar mass star.  The curves are measured in two bands: in Band 1, the planet's albedo is large and the moon's is small, and in Band 2, vice versa.  The curves are normalised so that the exoplanet phase curve peaks at unity. Bottom left: a normalised subtraction of the total flux in Band 1 from the total flux in Band 2 (after filtering to remove long period trends).  Bottom right: the same operation for a system where the moon orbits the star instead of the planet, with the same orbital period.\label{fig:reflective_planetmoon}}
\end{figure*}

The top row of Figure \ref{fig:reflective_planetmoon} shows the contributions from the phase curve due to the planet, and due to the moon, in both bands.  In Band 1, the exomoon contribution is virtually nil, and hence the total phase curve is indistinguishable from that of the exoplanet phase curve.  In Band 2, the albedo ratio $\alpha_s/\alpha_p=10^4$, which greatly exceeds the size ratio $R_p/R_s \sim 45$.  The total phase curve is now dominated by the exomoon, which we can see in the top right panel of Figure \ref{fig:reflective_planetmoon}. 

To isolate the periodic signal of the exomoon, we normalise the curve in each band $i$ by its mean:

\begin{equation}
\tilde{F}_i = \frac{F_i}{\bar{F}_i}
\end{equation}

We compute $\tilde{F}_1 -\tilde{F}_2$, and then pass this quantity through a bandpass filter to remove trends much longer than the lunar period (i.e. of order the planetary period), the results of which are displayed in the bottom left panel of Figure \ref{fig:reflective_planetmoon}.  We can perform this filtering with confidence, as we know the maximum permissible lunar orbital period from dynamical stability arguments (Figure \ref{fig:period_range}).  Note the amplitude modulation of the curve, which is characteristic of a planet-moon system.  If the moon had orbited the star, and not the planet, we would instead receive the steady-amplitude signal shown in the bottom right panel of Figure \ref{fig:reflective_planetmoon}.

We provide this example to illustrate the reflective limit of our calculations, but the effect strength is so small it is unlikely to ever be usable for detecting Galilean analogues, as we will see in the following section.  This technique may be more amenable to Earth-Moon analogues, where the satellite-planet size ratio is much higher, and the albedo ratio between the Moon and Earth is relatively high, given a judicious choice of wavelength.  This has already been demonstrated by \citet{Robinson2011} and \citet{Gomez-Leal2012} in detail, and we will show more general constraints in the following section.

\subsection{Detection Limits}

\noindent So what can we expect to observe in the highly reflective limit? If we specify a given moon-planet flux ratio $F_{s}/F_{p}$ that we believe observations can detect, then we can rearrange equation (\ref{eq:reflective_regime}) to obtain the detectable moon-planet size ratio:

\begin{equation}
\frac{R_s}{R_p} \geq \left(\frac{F_s/F_p}{\frac{\alpha_s}{\alpha_p} + \alpha_s\left(\frac{R_p}{a_{ps}}\right)^2}\right)^{1/2}.
\end{equation}

\noindent By considering detection limits in terms of $F_{s}/F_{p}$, we have implicitly assumed that the exoplanet phase curve is itself detectable.  In effect, we now ask: to what level of precision is the exoplanet phase curve known? Is this precision sufficient to yield a periodic signal indicative of a highly reflective exomoon? Conversely, we should also consider what levels of $F_{s}/F_{p}$ are insufficient to detect exomoons of a given radius or radius ratio.  We address the absolute detection limits considering current and future surveys, and sources of intrinsic phase curve variability in section \ref{sec:instruments}.

\begin{figure*}
\begin{center}$\begin{array}{cc}
\includegraphics[scale=0.4]{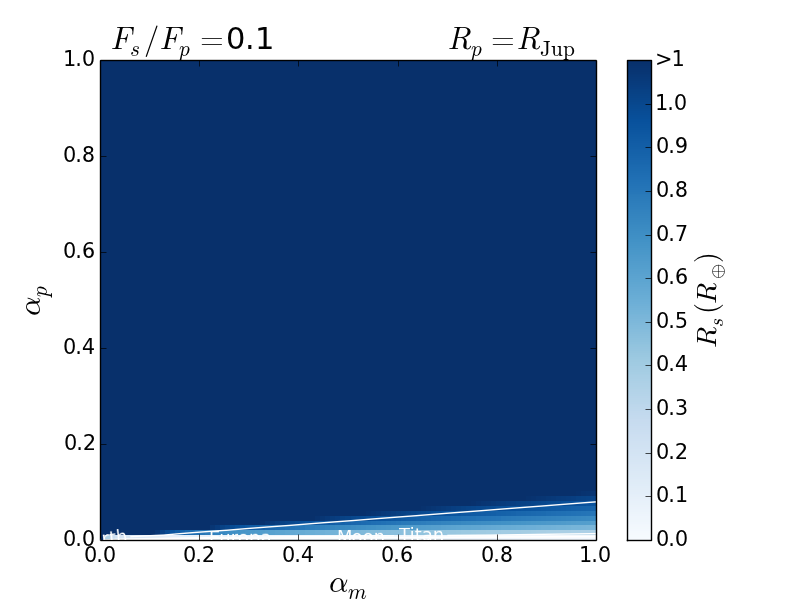} &
\includegraphics[scale=0.4]{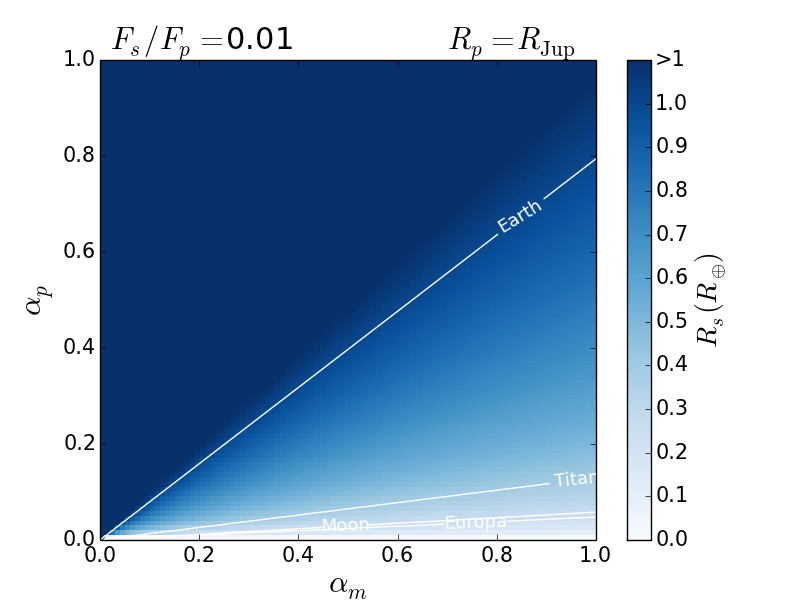} \\
\includegraphics[scale=0.4]{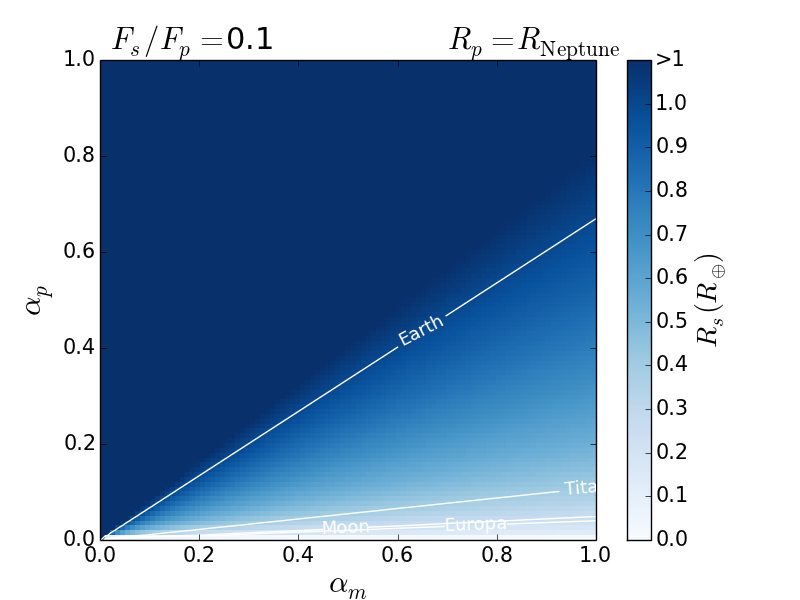} &
\includegraphics[scale=0.4]{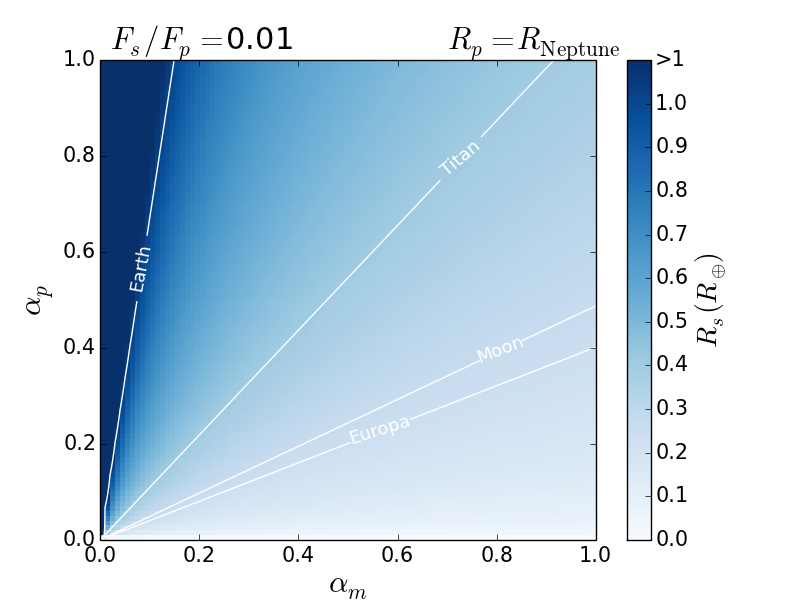} \\
\includegraphics[scale=0.4]{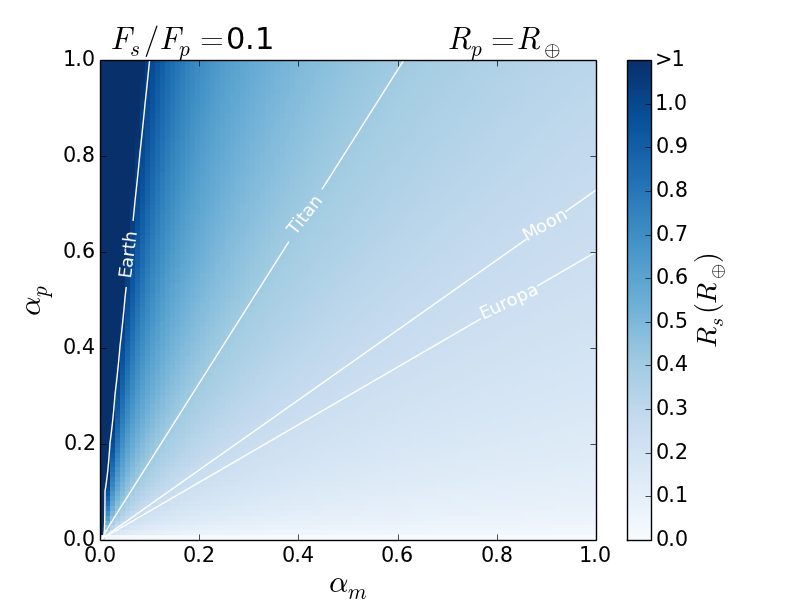} &
\includegraphics[scale=0.4]{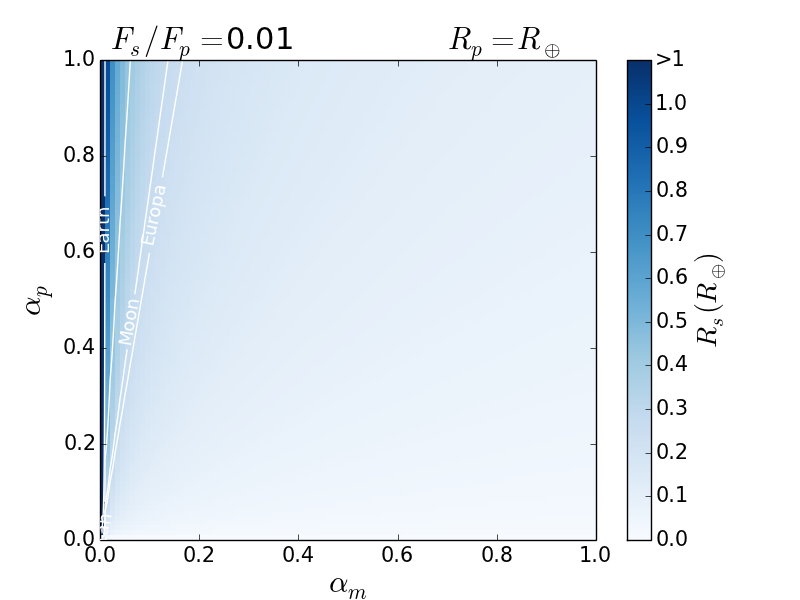} \\
\end{array}$
\end{center}
\caption{The detectable satellite radius ratio $R_s$ in the highly reflective regime where the temperature of both bodies is determined by the stellar radiation field, as a function of $\alpha_p$ and $\alpha_s$, for two different observing precisions: $F_{s}/F_{p}=0.1$ (left column) and $F_s/F_p = 0.01$ (right column), and three planet radii: Jupiter (top row), Neptune (middle row), Earth (bottom row).  The lunar semimajor axis is fixed at $a_{ps}=0.0045$ au. \label{fig:reflective_detectlimit}}
\end{figure*}

Figure \ref{fig:reflective_detectlimit} shows the detectable moon radius, as a function of $\alpha_p$ and $\alpha_s$ (assuming that we can characterise the phase curve to precisions of at least 10\% (left column) or  1\% (right column)).  Note that these limits are independent of $a_{p*}$, but they are ultimately not encouraging.

Being able to characterise an exoplanet phase curve with uncertainties below 1\% is a formidable challenge.  If we are only able to characterise the curve to within 10\%, then we are unlikely to detect moons below $R_s = 1\rearth$ orbiting a $1\rjup$ planet (unless we can identify wavelength regimes where $\alpha_s>0.5$ and $\alpha_p < 0.05$).  If the planet is Neptune-sized, then the limits on $\alpha_p$ are much less constraining, and Titan-sized moons come within grasp if $\alpha_s$ is large.  In the terrestrial regime a wide range of parameter space becomes available (although the measurement of the exoplanet phase curve itself becomes significantly more challenging).

We should therefore conclude that for the present time, moons that are not self-luminous are unlikely to be detected around giant planets using this method (at least for some time), although there may be promising regions of parameter space for Neptunes and super-Earths.  It is also worth noting that the $1.27 \rjup$ planet TrES-2b has an extremely low measured geometric albedo of $<2\%$ \citep{Kipping2011b}.  A highly reflective sub-Earth radius moon in this system would be detectable even with 10\% precision in the phase curve.

\subsection{The Highly Thermal Regime}

\begin{figure*}
\begin{center}$\begin{array}{cc}
\includegraphics[scale=0.4]{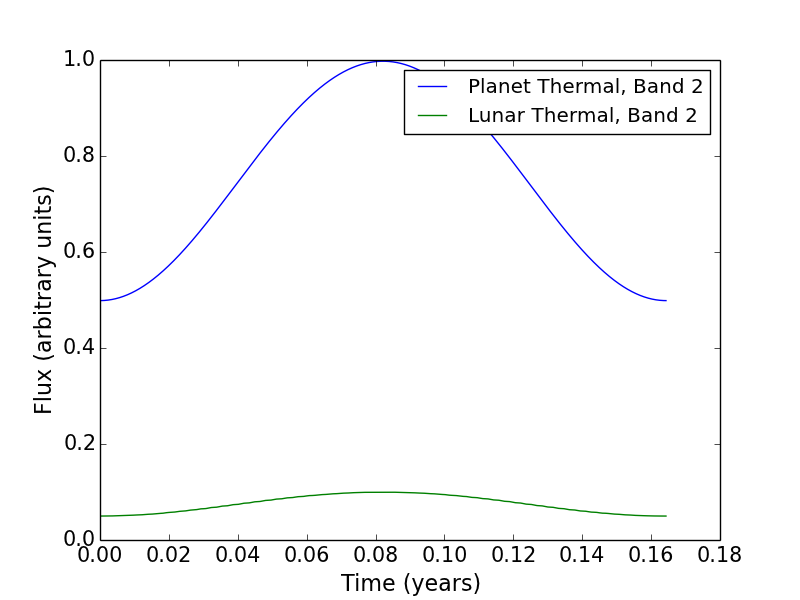} &
\includegraphics[scale=0.4]{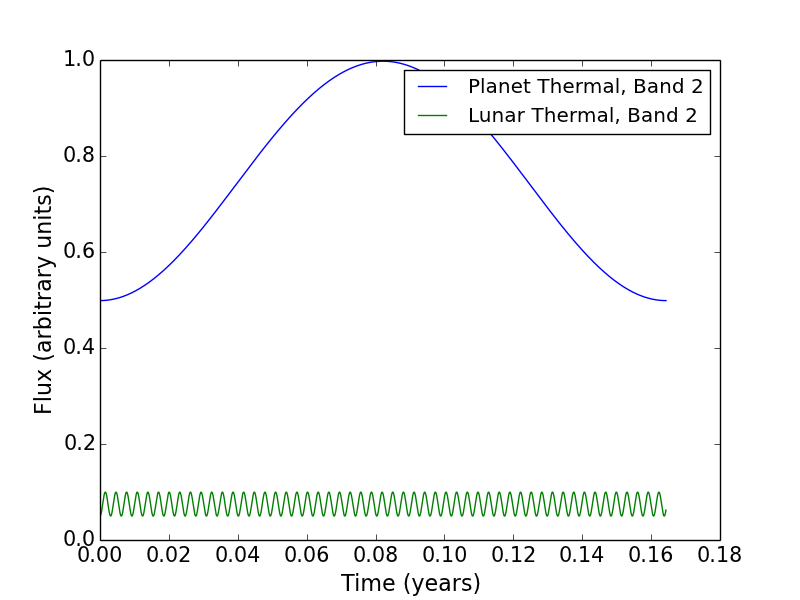} \\
\includegraphics[scale=0.4]{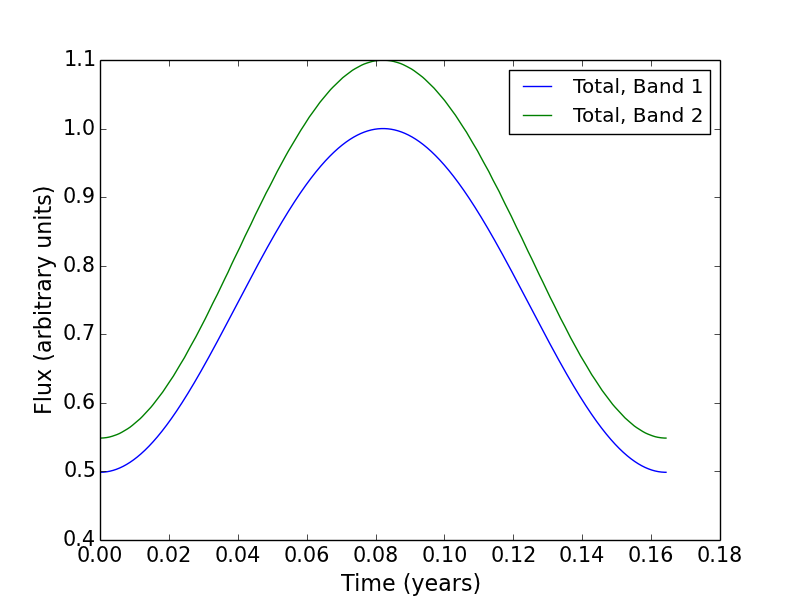} &
\includegraphics[scale=0.4]{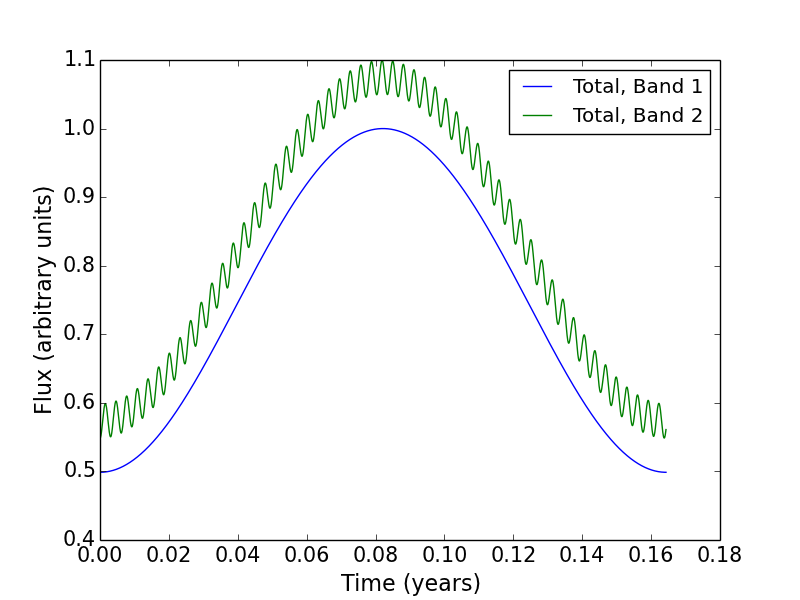} \\
\end{array}$
\end{center}
\caption{Isolating the exomoon contribution to the phase curve from a Jupiter-Europa system placed at $a_{p*}=0.3$ au from a solar mass star, in the highly thermal regime, where the lunar terminator is defined by the star (left column) or the planet (right column).  The curves are measured in two bands, and normalised so the peak exoplanet flux is unity.  In Band 1, the moon exhibits no thermal flux; in Band 2, the moon exhibits flux that is 10\% of the planetary thermal flux.  The planet's thermal flux is constant in both bands. Top: The individual contributions to the phase curve.  Bottom figures: the total flux in both bands.\label{fig:thermal_planetmoon}}
\end{figure*}

\begin{figure*}
\begin{center}$\begin{array}{c}
\includegraphics[scale=0.35]{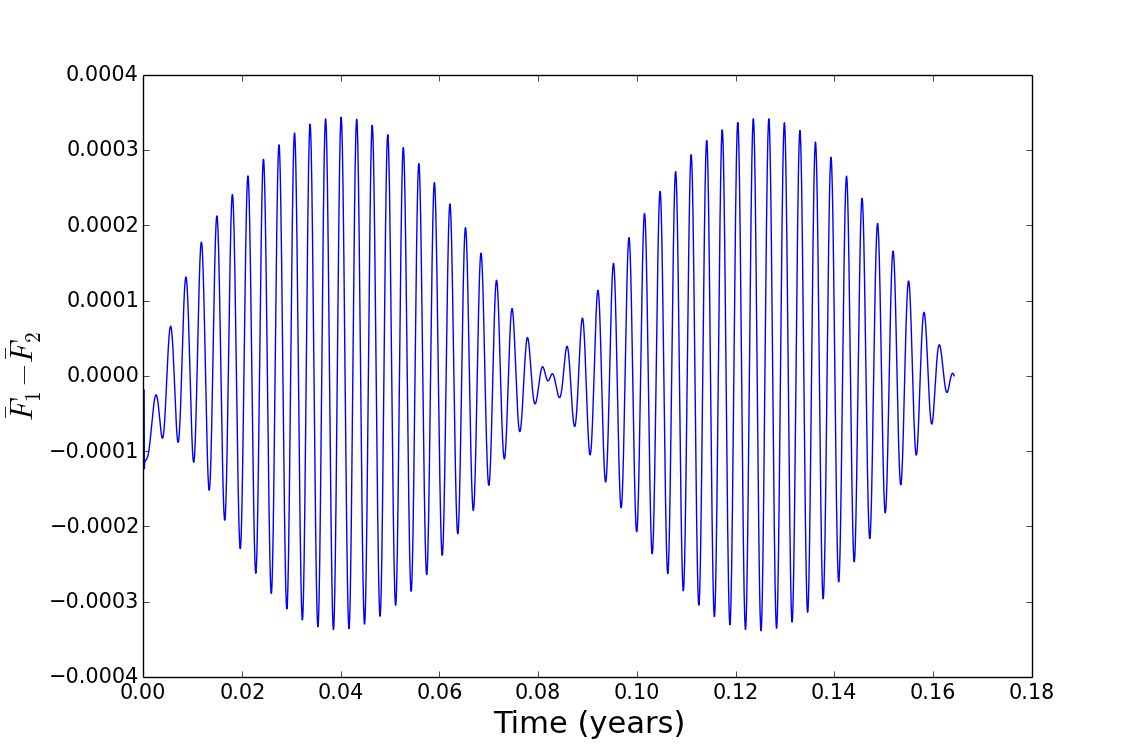} \\
\includegraphics[scale=0.4]{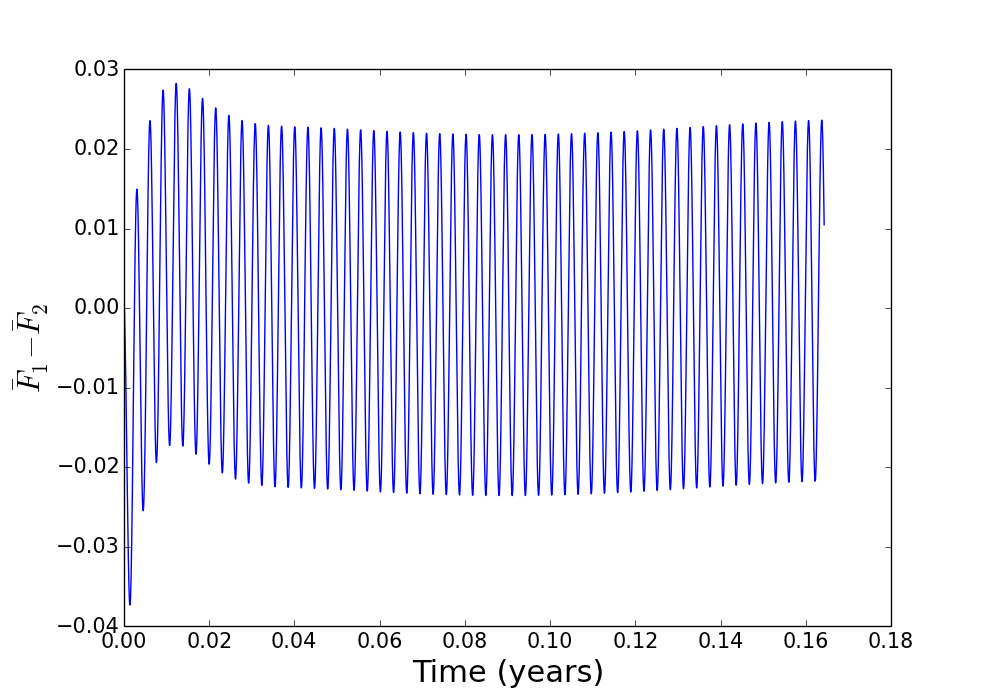} \\
\includegraphics[scale=0.4]{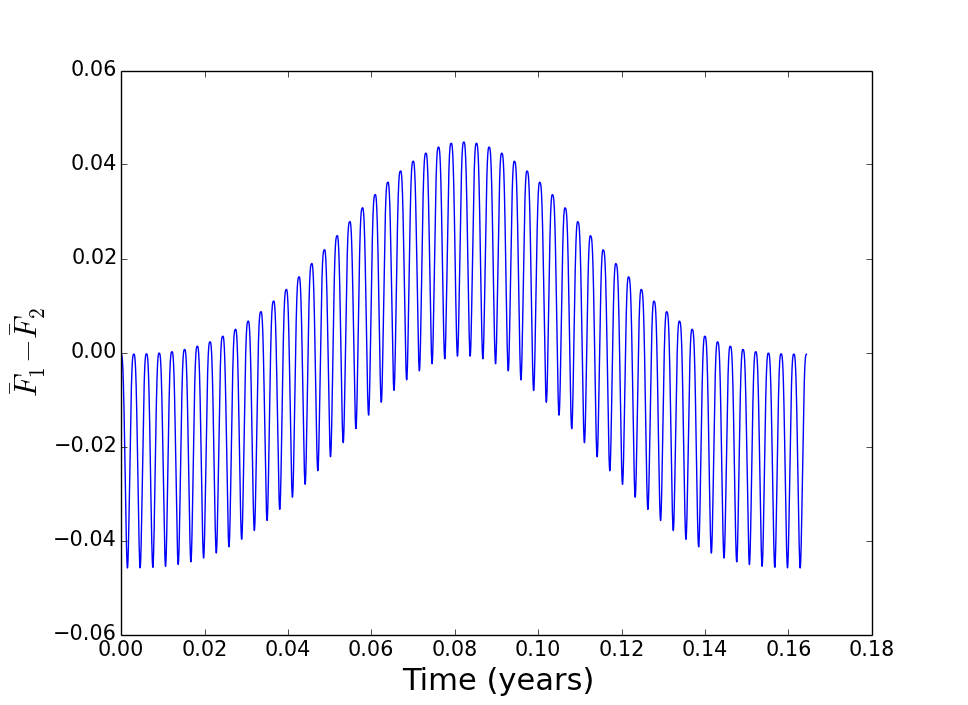}\\
\end{array}$
\end{center}
\caption{Normalised subtraction of the total flux in Band 1 from the total flux in Band 2 (after filtering to remove long period trends), for the highly thermal cases.  Top: where the lunar terminator is determined by the star.  Middle: where the lunar terminator is determined by the planet.  Bottom: the same operation for a system where the moon orbits the star instead of the planet, with the same orbital period. \label{fig:thermal_subtraction}}
\end{figure*}

\noindent We now investigate the nature of the highly thermal regime by reducing the albedo of both bodies in all bands to $10^{-4}$.  We set the thermal flux from the planet $F_{p,t}$ to be constant across both bands, and set the ratio of thermal fluxes from both bodies in each band to be $F_{s,t}/F_{p,t}=[0.0,0.1]$.  For both objects, we set the ratio of thermal flux from the day and night side $f_{night,p}=f_{night,s} = 0.5$.  Figure \ref{fig:thermal_planetmoon} shows the individual contributions to the phase curve generated by the planet and moon.  The left column indicates the curves generated if the lunar thermal flux is determined by the star, and the right column shows the case where the lunar thermal flux is determined by the planet.  We also show the normalised subtraction for both cases, and the equivalent subtraction if the moon was in fact orbiting the star (Figure \ref{fig:thermal_subtraction}).

If the star determines the lunar thermal flux, the exomoon curve's principal period is the planetary period, in much the same fashion as the reflective case.  The residuals derived from the normalised subtraction (top plot in Figure \ref{fig:thermal_subtraction}) are also extremely similar to those found for the reflective case, despite the change from a reflective Lambertian phase function to a dayside visibility function.

If the planet determines the lunar thermal flux, the exomoon curve's principal period is the lunar period, as we are most sensitive to the surface variation of flux as the moon moves in and out of view.  The combined phase-curve is a simple superposition of the exoplanet and exomoon periods, and the normalised subtraction (middle plot in Figure \ref{fig:thermal_subtraction}) is a simple sinusoid, which is easily distinguished from the case where both bodies orbit the star (right panel).  However, this is extremely similar to the two-planet residual in the highly reflective limit (Figure \ref{fig:reflective_planetmoon}).  Determining $\alpha_p$ would be required to break degeneracy in this case.

\subsection{Detection Limits}

\noindent As we did for the highly reflective regime, we can rearrange equation (\ref{eq:thermal_regime}) to obtain the minimum detectable size ratio as a function of the flux ratio and physical parameters, although we will also have to specify the planetary and lunar thermal fluxes to do so.  If we assume that both the planet and moon are blackbody emitters, then

\begin{equation}
\frac{F_{s,t}}{F_{p,t}} = \left(\frac{R_s}{R_p}\right)^2\left(\frac{B_\lambda(T_s)}{B_\lambda(T_p)}\right),
\end{equation}

\noindent Where $T_s,T_p$ are the effective temperatures of the two bodies.  The critical size ratio for detection depends on the peak wavelength of emission for the satellite:

\begin{equation}
\frac{R_s}{R_p} \geq \left(\frac{F_s/F_p}{\left(\frac{B_\lambda(T_s)}{B_\lambda(T_p)}\right) + \alpha_s\left(\frac{R_p}{a_{ps}}\right)^2}\right)^{1/2}.
\end{equation}

\noindent If we assume that both bodies are in thermal equilibrium with the stellar radiation field, then 

\begin{eqnarray}
T^4_p &= T_{\rm \odot}\left(1-A_p\right) \left(\frac{\rsol}{2a_{p*}}\right)^2 \\
T^4_s &= T_{\rm \odot}\left(1-A_s\right) \left(\frac{\rsol}{2a_{s*}}\right)^2.
\end{eqnarray}

\noindent where $A_p$ and $A_s$ are the Bond albedo of planet and satellite respectively.  As before, we approximate $a_{p*} \approx a_{s*}$, and hence

\begin{equation}
\frac{T^4_s}{T^4_p} = \frac{1-A_s}{1-A_p}.
\end{equation}

\noindent In other words, if the stellar radiation field dominates the lunar energy budget, it is likely that $T_s \approx T_p$ and hence the detectability of such moons is difficult (except if the planetary Bond albedo is unusually high and the lunar Bond albedo is unusually low).

\begin{figure*}
\begin{center}$\begin{array}{cc}
\includegraphics[scale=0.4]{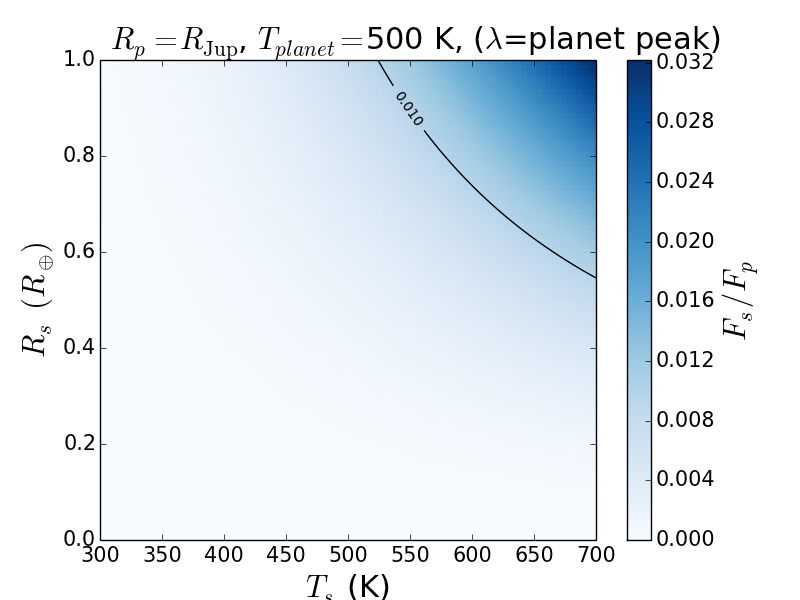} &
\includegraphics[scale=0.4]{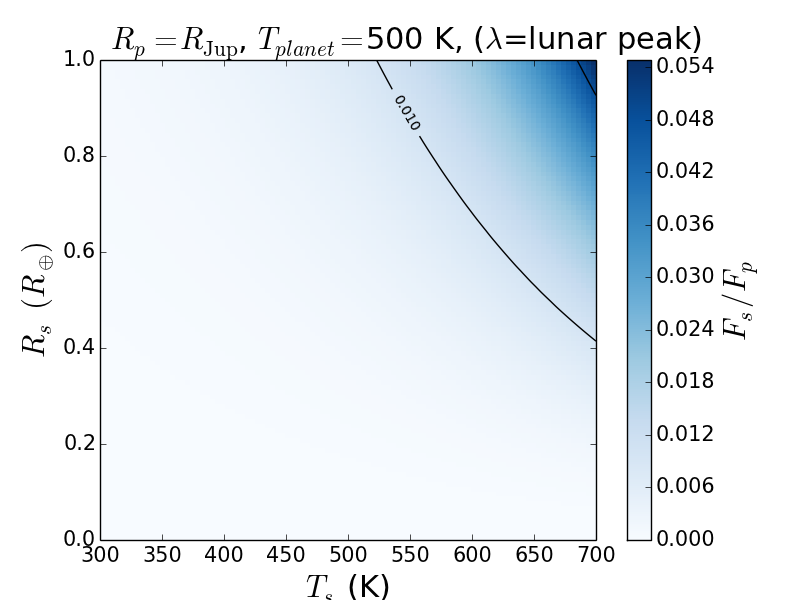} \\
\includegraphics[scale=0.4]{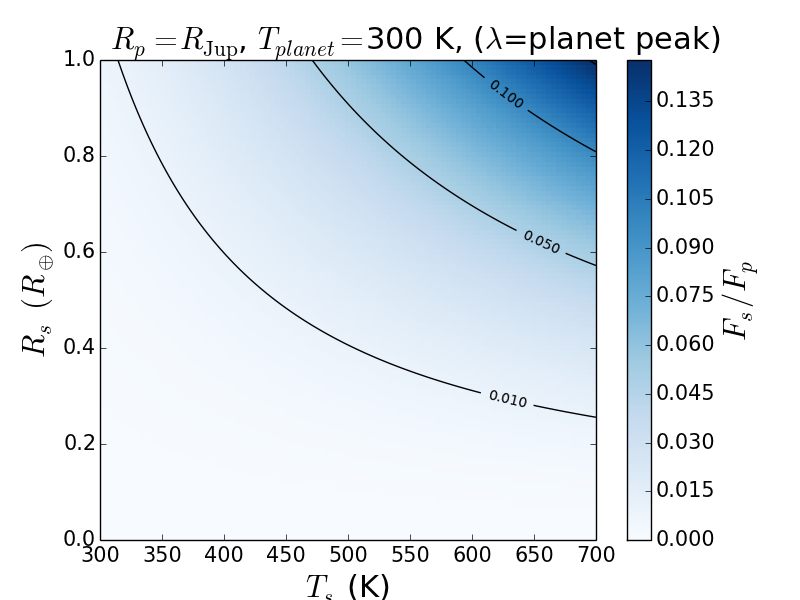} &
\includegraphics[scale=0.4]{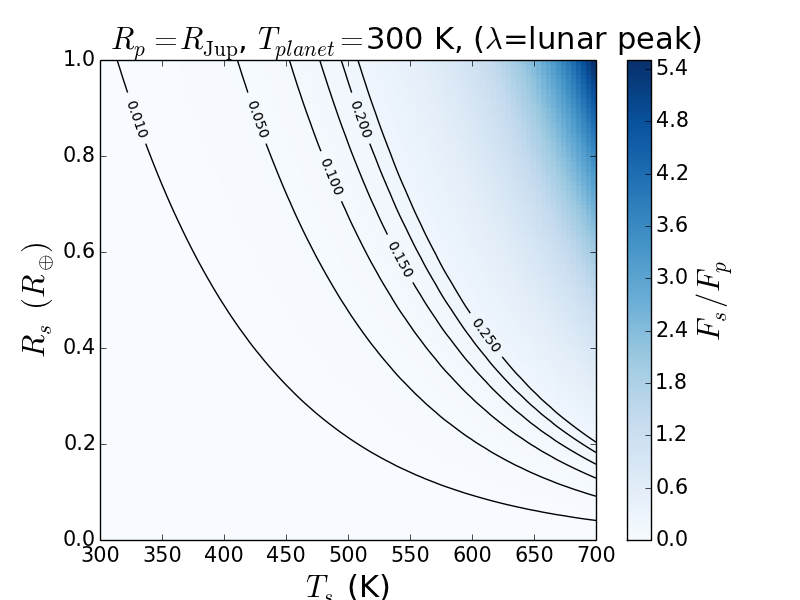} \\
\end{array}$
\end{center}
\caption{The detectability of tidally heated exomoons via PCSC.  The expected flux ratio $F_s/F_p$ is plotted (both in colour and in contours) as a function of satellite radius $R_s$ and satellite temperature $T_s$, given that $R_p=1\rjup$.  Plotted are the cases where $T_p=500$K (top) and $T_p=300$K (bottom).  Left and right column indicates which wavelength the flux ratio is measured at: either the peak of the planetary flux or the peak of the lunar flux respectively. Note the differing colour bar scales for each plot. \label{fig:thermal_detectlimit_jup}}
\end{figure*}

\begin{figure*}
\begin{center}$\begin{array}{cc}
\includegraphics[scale=0.4]{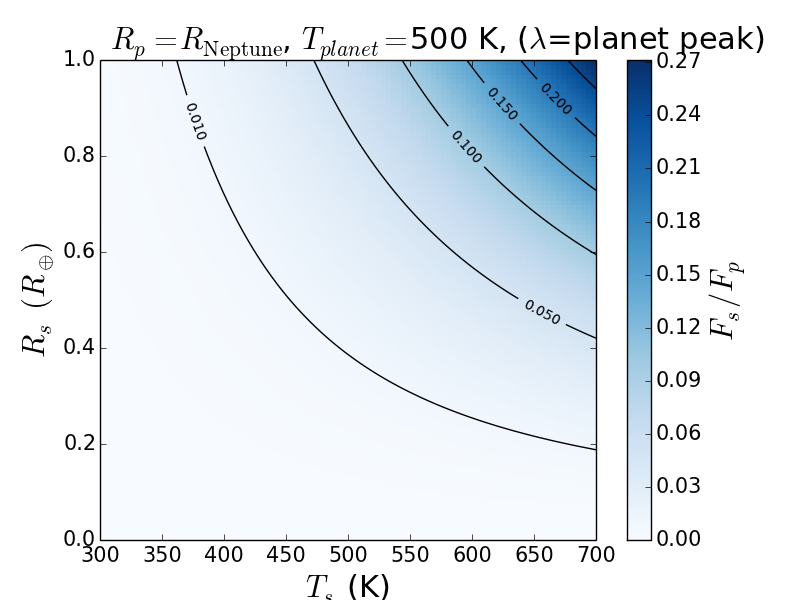} &
\includegraphics[scale=0.4]{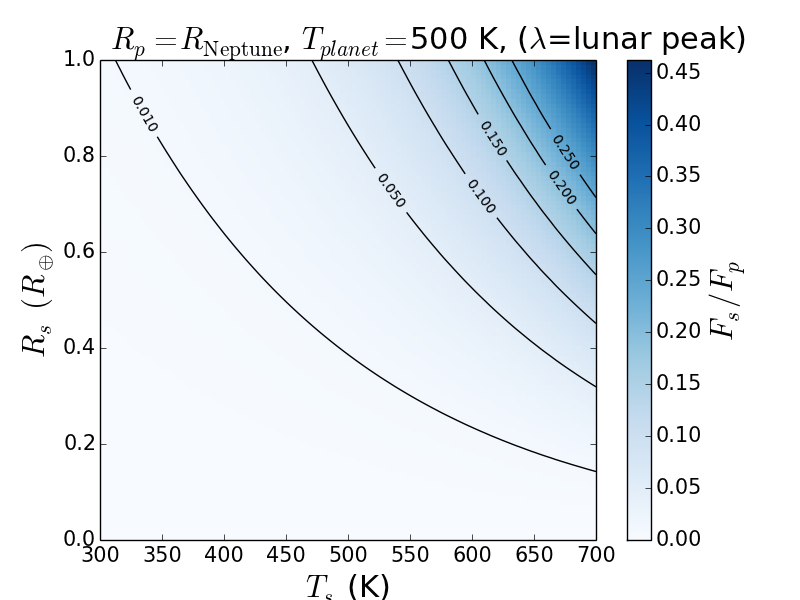} \\
\includegraphics[scale=0.4]{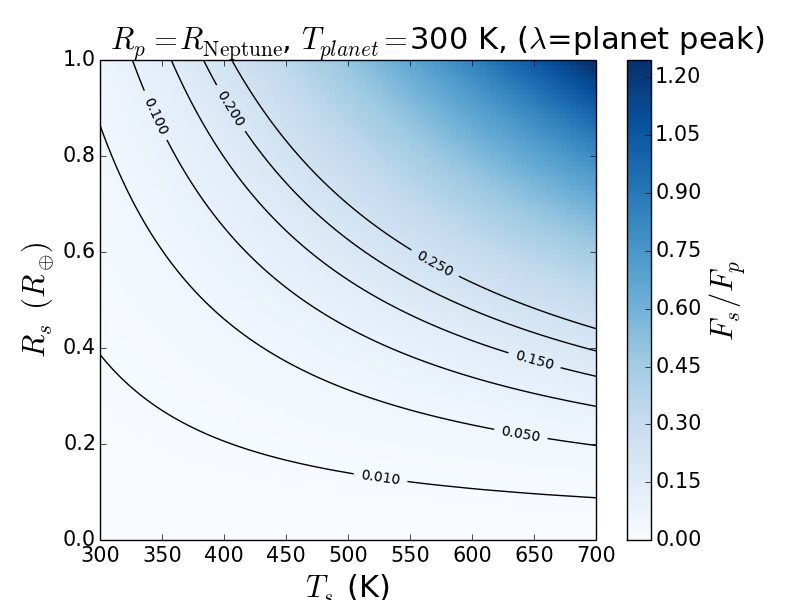} &
\includegraphics[scale=0.4]{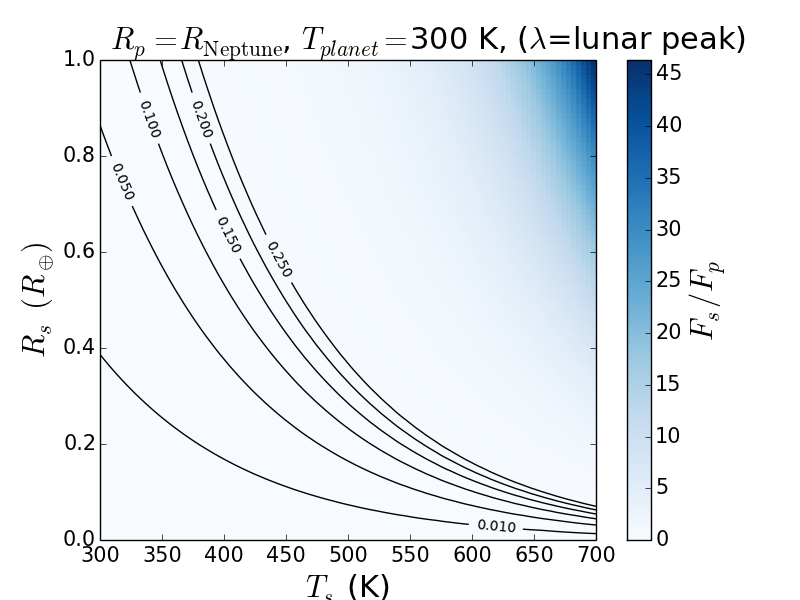} \\
\end{array}$
\end{center}
\caption{As Figure \ref{fig:thermal_detectlimit_jup}, but for a planet with radius equal to Neptune.  Again note the differing colour bar scales for each plot. \label{fig:thermal_detectlimit_nep}}
\end{figure*}

Of course, the temperature of the moon may not be determined by the stellar radiation field.  If the exomoon is strongly tidally heated, then it may well be that the moon surface temperature is significantly higher than the planet's regardless of albedo\footnote{This depends heavily on the redistribution of heat from the lunar dayside to the lunar nightside.  Even with strong tidal heating, effective heat redistribution will result in very low variations of thermal flux between dayside and nightside}.  Figure \ref{fig:thermal_detectlimit_jup} shows the detectability of tidally heated exomoons around a Jupiter-radius planet by phase curve spectral contrast as a function of moon temperature and $R_s$ (Figure \ref{fig:thermal_detectlimit_nep} shows the same for a Neptune-sized planet).  In both cases, the planets orbit an M star at $a_{p*}=0.2$ AU, and the lunar semimajor axis is fixed at $a_{ps}=0.0045$ AU (the resulting lunar detection limits are insensitive to both parameters, but the planet's detection is more likely for reduced $a_{p*}$).  We consider two different planetary temperatures: 300K and 500K.  

We also consider measurements made either at the wavelength corresponding to the Wien peak of the planet's blackbody emission, or the Wien peak of the moon, given $T_s$.  Measuring at the moon's peak clearly is best for detection of lunar photons, but measuring at the planet's peak is best for characterising the planetary phase curve in detail.  As we have seen in previous sections, comparing measurements of planet-dominated and lunar-dominated bands is the key to this detection method.

For Jupiter-sized planets at 500K ($\lambda_{peak}= 5.8\micron$), even extremely hot moons have a low contrast of $F_s/F_p < 0.1$ at the planetary peak.  Shifting to shorter wavelengths allows these hot moons to have a boosted contrast (top right plot of Figure \ref{fig:thermal_detectlimit_jup}).  For example, a 1$\rearth$ moon at $T_s=850 K$ would constitute around 25\% of the total phase curve signal.

If the planet has a temperature of 300 K, then the moon is more detectable even at wavelengths tuned to the planet's peak ($\lambda_{peak}= 9 \micron$).  At lunar-tuned wavelengths, the emission from a 850K moon dominates the phase curve signal, even for satellites as small as 0.2 $\rearth$!

Neptune-sized planets offer even better prospects, especially at $T_p=300K$.  A Titan-sized body only needs to achieve a temperature of around 500K to begin dominating the lunar-tuned band (while still achieving $F_s/F_p>0.05$ in the planet-tuned band).

What is less clear is the expected phase curve from tidally heated bodies. The temperature difference between day and night sides of a synchronously rotating, tidally heated moon is clearly a function of local rheology, and how tidal heat is redistributed from the interior.  

Finally, we should note that we have assumed a blackbody spectrum for the planet in this section.  A planet with strong absorption bands longward of a few $\micron$ would provide even better contrast to moons experiencing extreme tidal heat.

\subsection{Prospects for plausible $F_s/F_p$ measurements with current/future instruments \label{sec:instruments}}

\noindent To determine what values of $F_s/F_p$ we can expect to constrain in the future, let us begin with current calculations of the detectability of the phase curve of a single exoplanet in transit surveys.

Consider a transit survey with a duration given by baseline $B$, and cadence $C$.  If we are attempting to detect a signal with amplitude $A$ over one epoch of its period $P$, the signal-to-noise ratio is

\begin{equation}
SNR_1 = \frac{1}{\sqrt{2}}\frac{A\sqrt{\frac{P}{C}}}{\sigma}
\end{equation}

\noindent where $\sigma$ is the photometric precision.  We introduce a factor of $\sqrt{2}$ to accomodate the fact we are dealing with a sinusoidal signal rather than a step function.  If we are able to observe the maximum number $N$ epochs of the signal within $B$, then the combined signal-to-noise

\begin{equation}
SNR = SNR_1 \sqrt{\frac{B}{P}} =  \frac{1}{\sqrt{2}}\frac{A\sqrt{\frac{B}{C}}}{\sigma}
\end{equation}

\noindent If we are attempting to detect the phase curve induced by a single exoplanet (in the highly reflective regime), then 

\begin{equation}
A = \alpha_p \left(\frac{R_p}{a_{p*}}\right)^2 = 1.3 \,\rm{ppb}\, \alpha_p\left(\frac{R_p}{R_\oplus}\right)^2\left(\frac{a_p}{1 \rm{AU}}\right)^{-2}
\end{equation}

\noindent And hence the signal to noise ratio is

\begin{equation}
SNR = \frac{1.3 \,\rm{ppb}}{\sqrt{2}\sigma} \sqrt{\frac{B}{C}}\alpha_p\left(\frac{R_p}{R_\oplus}\right)^2\left(\frac{a_p}{1 \rm{AU}}\right)^{-2}
\end{equation}

\noindent We can immediately check the detectability with the \emph{Kepler} space telescope by substituting $B=1586.9\,\rm{d}$ (i.e. the entire Q0-Q17 baseline), $C=30 \,\rm{mins} = 1/48 \,\rm{d}$, and $\sigma=70 \,\rm{ppm}$ per six-hour timebin, where we have assumed our putative sample is in the upper tenth percentile of Kepler stars ($V=12 \,\rm{mag}$) to obtain this precision \citep{Christiansen2012}.  This gives

\begin{equation}
SNR = 5.3\times 10^{-3} \alpha_p\left(\frac{R_p}{R_\oplus}\right)^2\left(\frac{a_p}{1 \rm{AU}}\right)^{-2}
\end{equation}

\noindent As we can see, this limited sensitivity rules out Earth-sized planets possessing detectable phase curves, let alone any potential moons.  If we consider instead a Jupiter-sized planet orbiting at $a_p=0.2$ AU, then

\begin{equation}
SNR = 16.8 \alpha_p
\end{equation}

\noindent As $SNR$ is independent of $P$, we can obtain the $SNR$ of an exomoon phase curve by simply multiplying by $F_s/F_p$.  To maintain $SNR>5$, say, we are only able to probe signals of

\begin{equation}
\frac{F_s}{F_p} >0.29/\alpha_p
\end{equation}

\noindent If the exomoon's phase curve was generated purely by reflection, this would require $R_s >> R_\oplus$.  For extreme tidal heat, Earth-sized satellites may be detectable around warm Jupiters and Neptunes, but the satellite must have a surface temperature well in excess of 600 K (see the bottom right plots in Figures \ref{fig:thermal_detectlimit_jup} and \ref{fig:thermal_detectlimit_nep}).

This establishes the phase curve oscillation method as being of very limited use for current state-of-the art transit surveys.  We can now ask: what criteria must a transit survey satisfy in order for this technique to become feasible?

The cadence of future transit surveys (such as TESS, CHEOPS and PLATO), is typically of order 1 minute, although the photometric precision for TESS is likely to be worse \citep[see e.g. ][]{Rauer2014,Sullivan2015,Simon2015}.  

Let us now consider a putative survey where $C=1\, \rm{min}$, with a similar $B$ to Kepler.  We can rearrange to obtain the required photometric precision for detection with $SNR>5$:

\begin{equation}
\sigma < 128.8 \alpha_p \frac{F_s}{F_p}\,\rm{ppm}
\end{equation} 

\noindent For a relatively high amplitude exomoon signal of $F_s/F_p = 0.1$, then we must still demand a photometric precision of around 10 ppm, which is extremely difficult to achieve.  Failing this, we must then demand extremely long baselines $B$. 

It is also worth pointing out that our analysis has assumed white noise.  Correlated noise, instrumental systematics and the stellar activity inherent to the target will impose a floor on the possible precision.  This is significantly worse if the periodicity of the stellar activity is close to either the planetary or lunar periods.  We have also assumed that the planet itself does not show intrinsic brightness variations due to weather or other sources of atmospheric variability \citep[cf][]{Esteves2013,Webber2015,Armstrong2016a}.

On the other hand, binning of the data can help ameliorate systematics/activity, but this still requires long observing campaigns (and inappropriate bin sizes will destroy any lunar signal).  Binning also introduces an implicit dependence on the planetary period, as the effective improvement from binning a data train of fixed length is diminished for longer periods. 

If our moon is tidally heated, then it is possible that $F_s/F_p >>0.1$, but this is typically at wavelengths much longer than the typical visible/near IR bands used for exoplanet transit surveys, and will probably require space-based observatories.  The natural candidate is the James Webb Space Telescope \citep{Greene2016,Batalha2017}, and/or the future LUVOIR \citep{Bolcar2016} and HabEx missions \citep{Gaudi2017}. 

\section{Conclusions}
\label{sec:conclusions}

We have investigated the feasibility of analysing exoplanet phase curves to determine if there is an extra component produced by an exomoon, a technique we christen exomoon phase curve spectral contrast (PCSC).  This technique relies on multiple frequency measurements of the curve, and isolating frequencies where the moon is relatively bright compared to the planet, which can be the case at specific infrared frequencies where the planet's atmosphere is in absorption.

We find that in general, phase curve measurements will have to be extremely precise to detect even modest satellite-to-planet size ratios.  We calculate that for this technique to detect exomoons, a photometric precision of 10 ppm or smaller is required, while maintaining cadences of order 1 minute.  If the moon is strongly tidally heated, lower precisions may be sufficient, but will require observations at wavelengths greater than  $3-5 \micron$. 

This technique is similar to the spectro-astrometric method of \citet{Agol2015} in that both require measurements in multiple bands where different components of the system dominate, but is complementary in that \citet{Agol2015}'s method works for high-contrast direct imaging, not transits.  It also bears a resemblance to the mutual events detection method of \citet{Cabrera2007} - their method also uses direct imaging, but relies on observing a lunar transit or eclipse.  These events are quite short in time (of order a few hours), and hence very high cadence measurements are required over much longer campaigns, compared to PCSC.  Importantly, in contrast to both \citet{Agol2015} and \citet{Cabrera2007}, PCSC does not require individual targeting of sources, and can be directly applied to exoplanet transit survey data, given appropriate precision.

Exomoon detection for transiting exoplanets is perhaps more likely to proceed from e.g. transit timing and duration variation measurements (TTVs/TDVs), which typically rely on Bayesian inference to determine the moon's properties \citep[e.g.][]{Kipping2011}, and can probe satellite-to-planet mass ratios as low as $10^{-3}$ depending on the object \citep{Kipping2015}.  The additional information obtained from measuring the phase curve can assist in informing the priors that enter these calculations, and could reduce the permitted solution space for exomoon orbital and structural parameters, although not by much until photometric precisions improve.

As this technique is most suited to tidally heated exomoons, studying exoplanet phase curves for exomoon contributions will place stronger constraints on the existence of high temperature satellites of planets orbiting relatively cool stars.

\section*{Acknowledgements}

DHF gratefully acknowledges support from the ECOGAL project, grant agreement 291227, funded by the European Research Council under ERC-2011-ADG.  The author warmly thanks David Armstrong, Andrew Collier-Cameron, Ian Dobbs-Dixon, David Kipping, Joe Llama, Annelies Mortier and Hugh Osborn for stimulating discussions, and Tyler Robinson for an insightful review of this manuscript.  This  research  has  made  use  of  NASA's  Astrophysics  Data  System Bibliographic  Services.




\bibliographystyle{mnras} 
\bibliography{exomoon_PCO}







\bsp	
\label{lastpage}
\end{document}